# The Origins of the Highest Energy Particles in Nature – where we've got to and where we go next

**A A Watson**[1]

*School of Physics and Astronomy*
*University of Leeds, Leeds, LS16 5NU, UK*
*E-mail: a.a.watson@leeds.ac.uk*

In his Nobel Prize lecture Victor Hess urged that different instruments, working together, should be used to solve the problem of the origin of cosmic rays. I review some of the key developments that have opened up the new fields of direct and indirect multi-messenger astronomy and that are guiding us to the solution of this riddle. I then discuss, very briefly, some of the new instruments that are shortly to come on line and give examples to show the long lead-times from conception to implementation that occur in this field. I conclude with some remarks about very ambitious future projects. The paper is not intended as a review: rather it is an attempt to set down issues discussed in the Hess Memorial Public Lecture given at the 2019 ICRC in Madison, Wisconsin and accessible at https://www.icrc2019.org.

[1]A A Watson

## 1. Introduction

Hess's 1936 Nobel Lecture was entitled 'Unsolved problems in physics: Tasks for the immediate future in cosmic-ray studies'. The tasks set down included the discovery of the origin of cosmic rays, a topic to which he contributed when, by making one of his balloon ascents during a solar eclipse, he eliminated the sun as a possible source. Hess remarked that **"*to make further progress in the field of cosmic rays it will be necessary to operate all our resources and apparatus simultaneously and side-by-side……*".** During the intervening 80-plus years we have made significant advances in our understanding of the origin problem. The somewhat slow progress has not been because of a lack of imaginative ideas but rather because of the need to await technological developments and to develop the ability to attract adequate funding.

Below I try to show where we have reached in our search for the origin of the most energetic particles in Nature – the radiation that Hess detected was eventually shown to be charged particles with energies well beyond those found in radioactive sources or from the man-made accelerators of the time. Thanks to observations made during the last two decades or so, with a remarkable suite of instruments, it is safe to say that great strides have been taken towards discovering the origin of these extraordinary particles. Although the instruments have not been operated 'side-by-side', they have been operated simultaneously: this has opened the possibility of coincident observations of messengers from the Universe – a field now styled 'multi-messenger astronomy' - a term that also embraces the intimate connection between high-energy cosmic rays, neutrinos and photons. I also briefly review the next steps and give examples that show how long the lead-time often is between conception of an idea and a significant advance. Finally I will mention the potential of two very ambitious projects.

Neutrinos and electromagnetic radiation are generated through interactions of protons or heavier nuclei with matter and radiation. Key reactions are

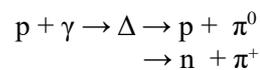

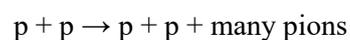

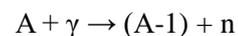

Charged pion decays yield neutrinos which travel through the Universe with minimal interactions retaining memory of their initial directions. This is also true for photons but only over smaller distances as interactions with ambient radiation fields produce lower energy photons and electrons. By contrast protons and heavier nuclei suffer small deflections in the magnetic fields lying between us and the sources so that the local direction of travel of a primary does not track-back closely to its source. Furthermore these deflections can cause the arrival of the particles to be delayed by hundreds of years with respect to electromagnetic or neutrino radiation. Interactions of the nuclei transform the composition of the accelerated material in a complex manner.

## 2. Results from some key instruments

### 2.1 Fermi Gamma-ray Satellite

Data from the Fermi gamma-ray satellite have been particularly useful in improving our knowledge of the gamma ray sky at energies of interest for understanding cosmic-ray origin. In particular, measurements of $\gamma$-rays above 1 GeV with the Large Area Telescope (LAT) have been used to make high-resolution maps of our Galaxy and to identify objects that are likely sources of cosmic rays at modest energies. More importantly, in the context of the highest-energy cosmic rays, has been the discovery of many powerful extragalactic γ-ray emitters, in particular blazars, together with a key measurement of the diffuse extragalactic background of photons at energies up to 100 GeV.

## 2.2 Ground-based gamma-ray detectors
Since the ground-breaking identification of the Crab Nebula as a source of TeV γ-rays thirty years ago, made using the Whipple Telescope to detect the flashes of Cherenkov light produced in the atmosphere [1], the field of ground-based gamma-ray astronomy has flourished. Over 200 sources, both galactic and extragalactic, have now been identified with the Whipple telescope and the second generation instruments (HESS, MAGIC and VERITAS). The γ-rays from these sources arise from synchrotron radiation from electrons gyrating in the ambient magnetic fields or from the decay of neutral pions. In the latter case these photons act as tracers of hadronic cosmic rays. For example, the HESS array of telescopes has been used to identify a hadronic Pevatron at the centre of our Galaxy [2]. With the MAGIC detector it has recently been possible to associate γ-rays of > 200 GeV from GRB 190114c [3].

Ground-based shower-detectors, used to detect the particles produced by high-energy γ-rays, have been installed in Mexico (HAWC) and in Tibet. Both arrays cover areas of ~20,000 $m^2$ and have the advantage of all-sky coverage 24 hours per day, though with poorer angular resolution. The HAWC detector comprises large tanks of water while the Tibetan array is of scintillators supplemented by an underground water-Cherenkov detector of 3,400 $m^2$. Amongst the many observations, a striking one is that the Crab Nebula produces γ-rays with energies well-above 100 TeV [4, 5].

## 2.3 Detectors of Astrophysical Neutrinos
The observation of extra-terrestrial neutrinos in the TeV – PeV has been achieved through the detection of muons and electrons that produce water-Cherenkov radiation in ice (IceCube at the South Pole) and in water (the ANTARES project in the Mediterranean and the Lake Baikal project in Siberia). These studies have opened a new astronomical window which promises to reveal the origin of the highest energy cosmic rays. The most convincing evidence for extra-terrestrial neutrinos is from the IceCube Collaboration who have reported a signal significant at the level of 5.7 sigma [6]. Additionally, as discussed below, observations with the IceCube detector have initiated the field of direct multi-messenger astronomy.

## 2.4 Detectors of the Highest-Energy Cosmic Rays
The Pierre Auger Observatory located in Western Argentina and the Telescope Array sited in Utah, USA, are dedicated to the study of the highest-energy particles. The Telescope Array comprises 507 scintillation detectors spread over 700 $km^2$ while the Auger Observatory has 1600 water-Cherenkov detectors deployed across 3000 $km^2$. These devices have been operated for 10 and 15 years respectively: both use fluorescence detectors to improve event reconstruction and to determine the energy scale. Measurements of the energy spectrum from the two instruments are in agreement on the form of the major features in the spectrum while a combined study of the arrival directions at the highest energies has revealed a weak correlation of regions of higher than average intensity with the super-galactic plane. The longer operating time of the larger Auger instrument has revealed a dipole anisotropy of (6.6 +1.2/-0.8%) that increases with energy [7] and has enabled the mass of the primaries to be estimated up to ~70 EeV.

The water-Cherenkov detectors of the Auger Observatory are 1.2 m deep and have proved extremely useful in searches for the ultra-high energy neutrinos expected to reveal themselves at large zenith angles through the presence of showers with the characteristics of a hadronic-initiated event from a more vertical direction. A remarkable shower displayed, but most likely produced by a hadronic primary, is shown in figure 1. So far no neutrino events have been identified but limits set on the flux of diffuse neutrinos are important in the context of discovering the origin of cosmic rays as they help constrain the mass composition [8].

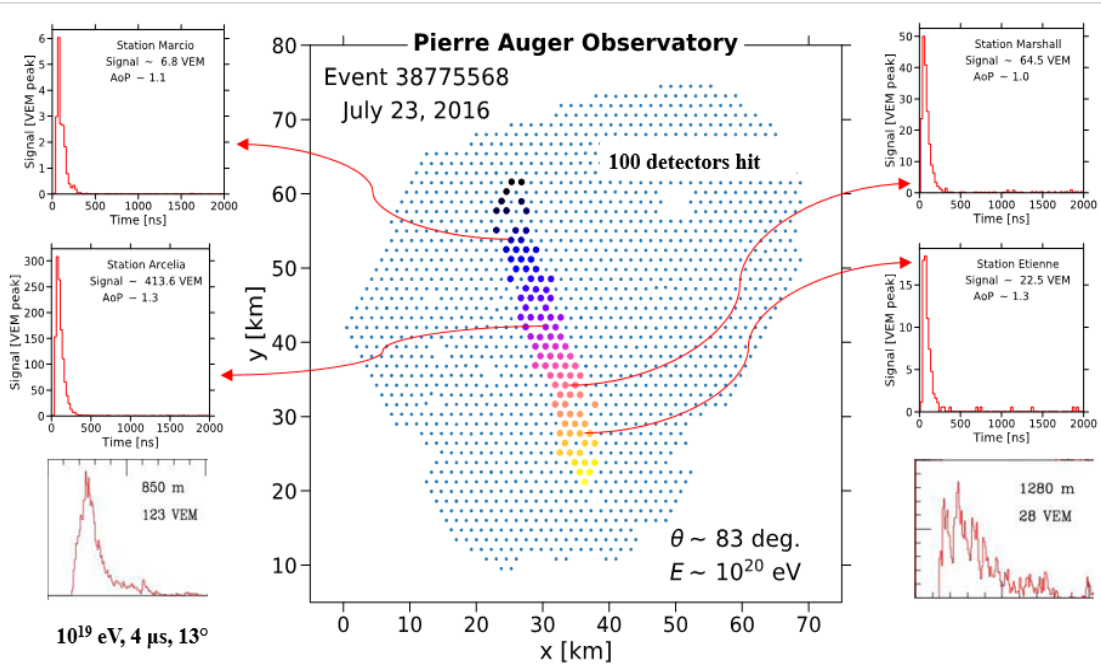

**Figure 1:** A shower of very large zenith angle in which 100 of the water-Cherenkov detectors of the Auger Observatory were struck. The FADC traces recorded at four of the detectors are shown and are very short. Two other, much longer, traces from a near-vertical event are also shown: a neutrino arriving at a large zenith angle is expected to show similarly long FADC traces.

## 3.    Multi-Messenger Observations

### 3.1 Direct Multi-messenger Observations

On September 17th 2017 the IceCube Collaboration issued an alert reporting the arrival direction of a muon-neutrino of ~290 TeV. It came from within 0.1° of a blazar (TXS 0506 + 056) at a distance of 1.7 Gpc. The blazar had been observed by the Fermi LAT to be in a flaring state since April of that year and was also identified as an emitter of $\gamma$-rays up to 200 GeV using the MAGIC telescope. The significance of these coincidences arising by chance was estimated as 1/1000. Subsequently the IceCube Collaboration searched their database for other neutrinos from this direction resulting in the discovery of a 150-day period during 2014/2015 when 15 neutrinos were detected against a background expectation of ~6 events. Details of these observations are summarized in figure 2 [9].

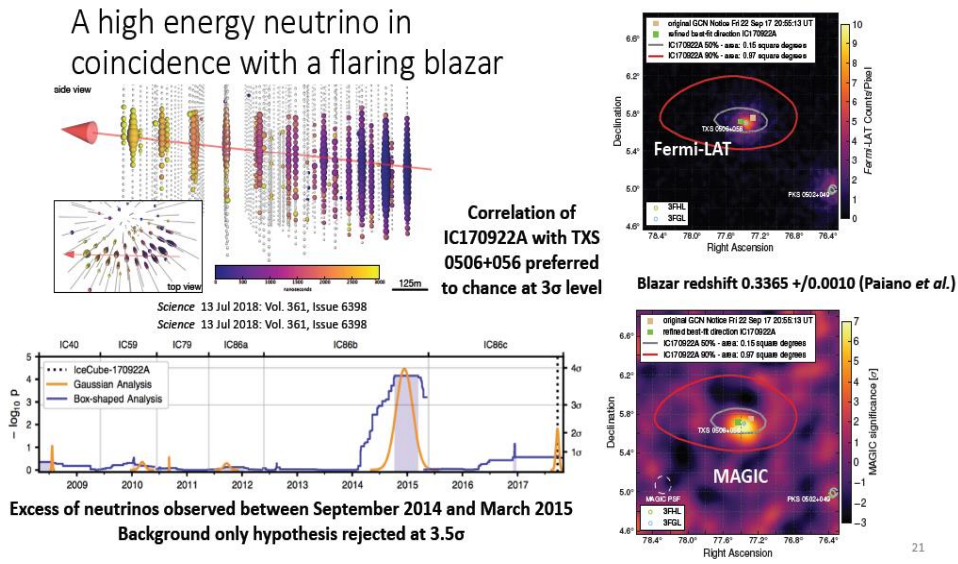

**Figure 2:** Data associated with neutrinos and γ-ray signals from TXS 0506 + 056 [9].

It is known from the work of the Auger Collaboration that cosmic rays above 8 EeV come from sources outside our Galaxy [10] so this first multi-messenger detection identifies blazars as one of the possible sources of the highest-energy cosmic rays.

Phenomenological analysis of details of the event indicate that TXS 0506 + 056 may not be a typical blazar and it has been speculated by Halzen [11], following a suggestion by Biermann, that this particular object might have arisen from the merger of two galaxies.

### 3.2 Indirect Multi-Messenger Observations

The connection between the highest-energy cosmic rays, neutrinos and the background flux of ~100 GeV γ-rays was pointed out long ago, with the striking observation that the energy flux of the three different components is roughly similar. The degradation of the energy in the UHECRs is like trickle-down economics, with the massive energy flux in the UHECR being shared between lower-energy cosmic rays, neutrinos and photons via the well-known processes listed in the introduction. Of the several analyses of this situation that have been made I find that discussed in [12, 13] particularly compelling for two reasons. Firstly, this study makes use of the detailed measurements of mass composition derived from the observations with the Auger Observatory and, secondly, the hypothesis offers a natural explanation for the absence of any anisotropy around 1 EeV, where the present limit is below ~1% and where protons have been shown to dominate the cosmic-ray beam. A light component from extra-galactic sources, below the ankle, gives a natural explanation for the absence of any significant anisotropy. A further distinctive feature of this work is that different hadronic models are considered, thus giving some idea of the systematic uncertainties associated with the predictions, a feature addressed by too few phenomenologists.

The model discussed in [12, 13] leads to an explanation of the cosmic-ray energy spectrum and the complex evolution of the primary-mass composition with energy caused by the photo-disintegration of nuclei in the photon field surrounding the acceleration region. The light composition below the ankle is explained as is the mixed composition above it. This model for the sources does not require either a detailed description of the magnetic field or of the size of the region. What is important is the ratio of the interaction time of the particles in the photon field to the escape time, the energy-dependence of which is modelled as a power law in rigidity.

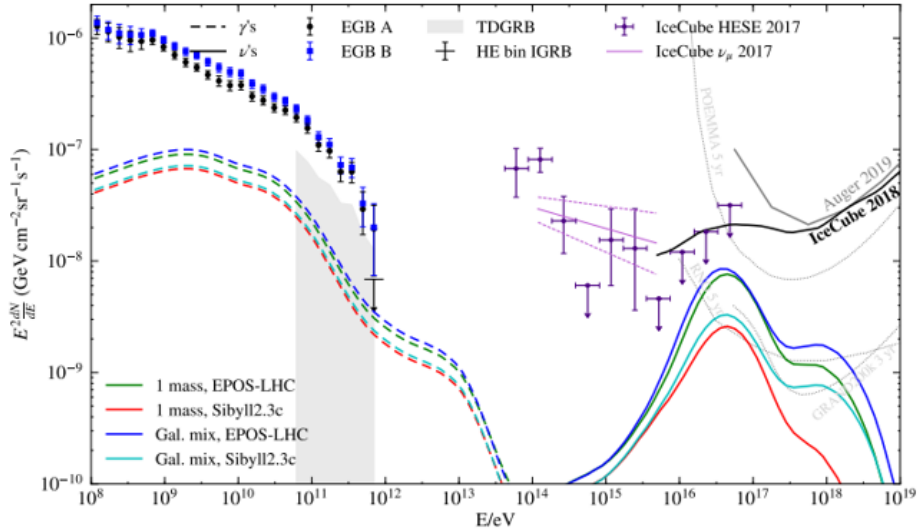

**Figure 3:** Adaption of a figure from [13] showing how the energy flux is shared across the high-energy cosmic ray, neutrino and γ-ray phase space.

The results shown in figure 3 are adapted from [13]. The requirements to match the observations of mass composition and energy spectrum reported by the Auger Collaboration lead to limits for the neutrino fluxes above 5 PeV and for the background γ-ray flux at ~100 GeV consistent with observation.

### 3.3 Conclusions from Multi-messenger observations
The direct observation of a neutrino from TXS 0506 + 056 and the availability of models for sources of high-energy cosmic rays that make predictions in agreement with observation is very encouraging. An unusual blazar has been identified as the source of some IceCube neutrinos [9, 11], while particles from a class of objects in which either a narrow range of nuclear masses or a composition similar to that found in the Galaxy, are accelerated and can explain the Auger data [13].

### 4.     The next generation of detectors
Impressive progress towards the solution of the problem of the origin of cosmic rays can be anticipated from instruments that are being brought into operation or are at advanced stages of planning.

A third generation device, the Cherenkov Telescope Array (CTA), being built by a global collaboration, will have imaging telescopes of three different sizes to explore the range from 10s of GeV to 300 TeV. These are planned for deployment on La Palma and at the ESO site in the Atacama Desert. The intention is to operate CTA as an open observatory. The wide range of science to be addressed includes searches for dark matter particles, deviations from Einstein's theory of special relativity, in addition to tackling more conventional cosmic ray and astronomical problems.

Detectors designed to give all-sky coverage of γ-ray sources in the Southern Hemisphere, as HAWC does for the north, are being planned. These, SWGO [37] and ALTO, will be air-shower arrays planned to have features effective for discriminating between hadronic and γ-ray primaries with high efficiency.

A particularly impressive new effort is the LHAASO detector in China which has begun to take data to study cosmic rays and γ-rays in the decades around 1 PeV. 1800 detectors are deployed over 22,500 m² at an altitude of 4410 m. Data from this device will contribute in a major way to our understanding of cosmic-ray origin and γ-ray emission in the medium energy region over the next decade [14].

To enhance observations of astrophysical neutrinos, there is progress on three fronts. In Lake Baikal, a gigaton detector (GVD) is under construction to cover the energy range from TeV to 100 PeV [15]. The IceCube collaboration is planning GEN2 which will have a volume of 10 km³ and an angular resolution of 0.1° [16], while in the Mediterranean the KM3NeT collaboration are constructing two detectors [17]. One, off the coast of France, is planned to have 115 photomultiplier modules and will concentrate first on neutrino physics using atmospheric neutrinos, while the second detector off the coast of Italy, with 230 modules, will be more focused on astrophysics. The detectors are expected to start taking data in 2024 and 2026 respectively.

For the study of the highest energies both the Auger and Telescope Array Collaborations are moving forward with ambitious plans. The Telescope Array will expand the area of their array by a factor of four, making it nearly as large as the Auger Observatory, with the primary physics goal of studying the 'hot spot' identified by them some years ago in more detail [18]. The Auger Collaboration are deploying 4 m² scintillators on top of each of the 1600 water-Cherenkov detectors and in addition are adding radio antenna at every station. The goal is to identify the mass of primary particles on an event-by-event basis with the objective of enhancing anisotropy studies [7].

Results from these devices will surely form much of the material to be discussed at the next 5 to 10 ICRC meetings.

## 5.     The lead time to develop and exploit technologies

So the fields of study that are important for understanding the origin of the highest energy cosmic rays are all in good shape with natural developments of existing techniques now properly funded. However the time between the conception of an idea, or of the implementation of new instrumentation, or of some novel methods that benefit analysis, can be strikingly long. I next want to give some examples, partly as an explanation for why progress towards the solution of the origin problem has been so slow. Several such topics were discussed at the first ICRC that I attended in London in 1965.

It is hard to imagine how our field could have advanced without the availability of the Monte Carlo technique but it took around 20 years from the invention of the method by Ulam and von Neumann for sufficient computer power to be widely available so that calculations relevant to air showers were possible. Results from such work were first reported at the London meeting: those carried out by Hillas were made on a machine at Leeds University having only 64 k of memory, a KDF9 which weighed about 5 tonnes. The technology has advanced!

The ideas of Artificial Intelligence (AI) and Neural Networks (NN), now gaining wide use across our field, have had an even longer gestation. I am a novice in this area but my reading suggests that the early claims made for the power of these concepts were often over-hyped. Historians speak of two 'AI winters', 1974-1980 and 1987-1993, during which funding for AI research largely dried up. In 1991, for example, massive Japanese investment was curtailed when objectives such as using computers to have a casual conversation had not been met, a target only achieved in 2010. During 1991 around 300 commercial AI companies were shut down or went bankrupt.

Interestingly, during the second winter, AI was being used to address problems of importance to the astroparticle-physics community. Halzen, Vazquez and Zas [19] and Reynolds [20], developed NN-techniques to analyse images in TeV gamma-ray telescopes. Jay Perrett tells me that he tried to interest Michael Hillas in the idea around this time but, for reasons not known to me, Hillas chose not pursue this line of attack. Perrett and van Stekelenborg, part of the team working in the Bartol Institute on analysis of data from the SPASE experiment at the South Pole, used the NN technique to improve the speed of shower reconstruction [21]. Their work was published but a most interesting prediction on the next sunspot cycle, made in 1988 and which proved to be accurate, was rejected by Nature. Using NN methods they forecast that the peak of the next solar maximum would occur in 1989 whereas experts had predicted 1991/2. Furthermore these experts had expected a monthly average of around 300 sunspots rather than the 212.5 observed and the 219 ± 11 predicted using the NN method. The data from the rejected paper are shown in figure 4.

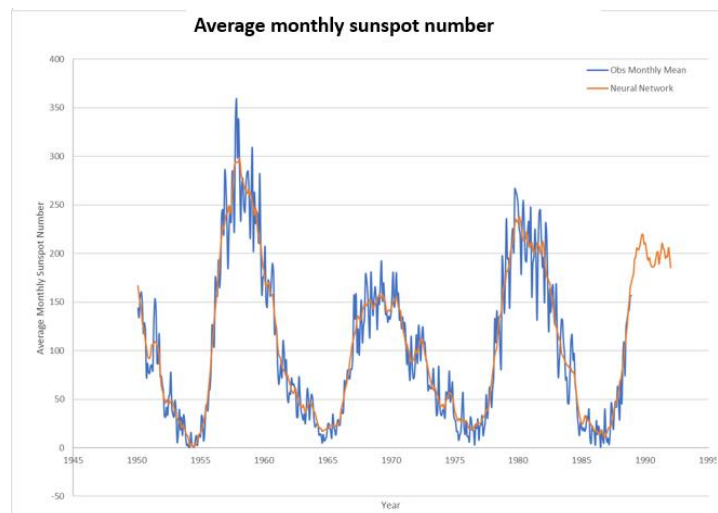

**Figure 4:** The number of sunspots and the time of sunspot maximum as predicted by Perrett and van Stekelenborg in 1988 (unpublished plot from J C Perrett).

During the London Conference, Greisen gave an invited paper ('Highlights in Air Showers 1965') which was insightful and influential and is still well-worth reading [22]. Amongst the topics he discussed was the detection of air showers using the radio method which had just been demonstrated [23]. Greisen commented '*The technique is barely in its infancy…. We feel confident it is a significant breakthrough and further study will reveal ways of obtaining information about showers not available by other means'*. His prediction would eventually turn out be correct but it was to be an infancy lasting nearly 40 years as the method fell into disuse around 1975, largely because of the inability, at the time, to monitor the geo-electric field adequately. It was revived early in the 21st century and is now a key tool being exploited at LOFAR to study the depth of shower maximum and at the Auger Observatory as noted above. It is also being exploited in several projects in Antarctica: through flights of ANITA to look for signals from neutrinos produced in the Antarctic ice, in the IceTop project to improve energy estimates, and to carry out searches for neutrino signals in Antarctic ice, an idea first proposed by Zas, Stanev and Halzen in 1992 [24], through the Askaryan Radio array and the Radio

Neutrino Observatory.  The radio method also underpins the ambitious GRAND neutrino detector of 200,000 km$^2$

A further technique that Greisen promoted was the fluorescence method.  It is unclear who first had the idea of using this auroral-like light to make calorimetric estimates of the energy of the primary particle initiating an air shower.  The method was first discussed in Japan in the mid-1950s in a report from the Institute of Nuclear Studies and was brought to the attention of a wider audience by Suga during a conference in La Paz in 1962 [25].  Chudakov mentioned his own work on this topic in the discussion following Suga's presentation though Greisen did not speak of his program that was already underway at Cornell.  First detections made in Japan were reported in 1969 [26] but it was not until 1977 that a prototype of what was to become the Fly's Eye instrument was tested by the Utah group at the Volcano Ranch array [27] with the first physics results following in 1985, 28 years after the first Japanese discussions.  The Japanese work, and similar efforts by Greisen and his group, were terminated largely because of the climatic conditions at the respective sites.  The huge advantages of a fluorescence detector operating in temporal coincidence with a shower array was demonstrated at the Auger Observatory in 2001.  This 'hybrid method', now also used at the Telescope Array, greatly improves the accuracy of reconstruction of shower events.

At the Jaipur ICRC in 1963, the Sydney group, led by C B A McCusker, had described a method of operating air shower arrays without the need to connect the detectors by cables [28].  This brilliant innovation of 'stand-alone detectors', conceived by Murray Winn, led to the construction of an array of ~100 km$^2$ at Narribri in New South Wales, Australia.  I have a vivid recollection of Greisen, during his invited talk, asking McCusker how large the Sydney array would be.  McCusker's reply, '*How big is Australia*?', has always stuck in my mind as, in principle, Winn's technique does allow for the construction of an array as large as one could fund.  Unfortunately, execution of the method demanded technology on the edge of that available in the late 1960s, making operation of the array difficult, and it was not until 2001 that the Auger Collaboration succeeded in implementing the Sydney ideas with high efficiency.  To achieve this, GPS for inter-detector timing, methods from cellular telephone technology to collect data across large areas and solar power to drive the electronics, were all needed.  Here the delay was nearly 40 years.

In Greisen's presentation mention was also made of Fazio's plan to build the Whipple telescope which was to lead to the discovery of the Crab Nebula as a TeV emitter some 24 years later.

In London the first results from the experiments in the Witwatersrand and Kolar Gold field mines, targeted at searching for neutrinos produced in the atmosphere, were described.  Five years earlier Markov had proposed that searches for astrophysical neutrinos might be made by installing detectors in a lake or a sea to be shielded from cosmic rays, determining the direction of neutrino-induced charged particles with the help of Cherenkov radiation [29].  This idea was advanced through the Dumand project, the efforts in Lake Baikal, and, later, the ANTARES and IceCube programs.  The IceCube device, the most successful of these efforts, was completed in 2011.

Thus the time needed to produce effective cosmic-ray instruments is rather long and, together with the historic shortage of funding that has blighted the field until recently, we can have some appreciation as to why progress in answering the question of the origin has been so slow.

## 6.      Ambitious plans for the future

In my view, the most ambitious plan is the one conceived by John Linsley in the late 1970s. His vision was to deploy a fluorescence detector in space. Some idea of Linsley's style can be gathered from his hand-written note seen in figure 5. He liked to refer to work in the air-shower region as the study of 'Auger Radiation', a homage to one of the pioneers in the study of this phenomenon and who had nominated him for the Nobel Prize in 1980 and 1981.

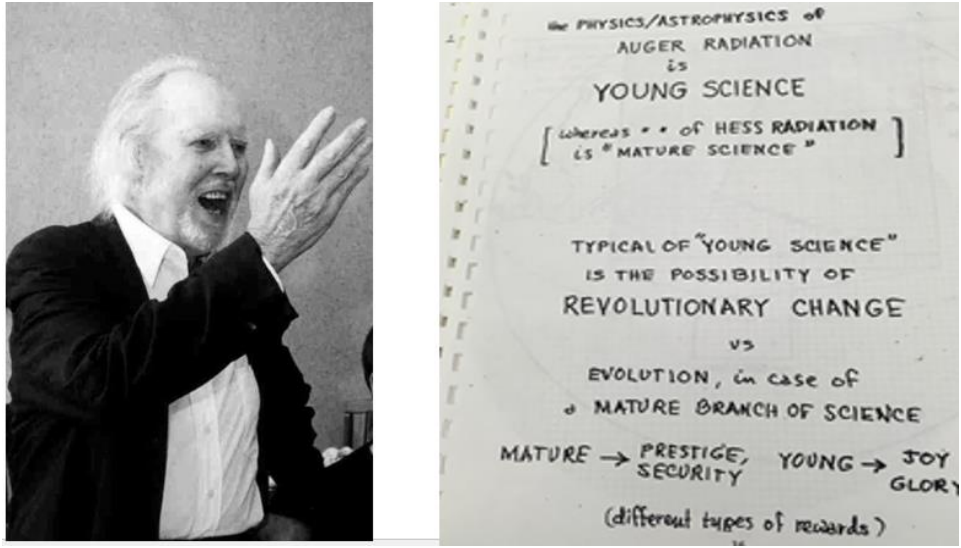

Figure 5: Left: John Linsley in the mid-1990s (photo credit: Mikhael Panasyuk); Right: a page from Linsley's notebook (FermiLab Archives).

A proposal to ESA to fly a single fluorescence detector was developed by Linsley, with L Scarsi and Y Takahashi as collaborators, with the name EUSO (Extreme Universe Space Observatory). Despite an impressive phase A study, ESA funding problems led to the project being dropped, but later to be picked up as JEM-EUSO with plans to mount such a device in the Japanese module on the International Space Station. This plan also foundered but the EUSO concept remains very much alive with a mini-EUSO now docked on the Space Station and observations scheduled to start with the arrival of trained astronauts in October 2019. This mission is one of several that are part of the program behind the POEMMA (Probe of Extreme Multi-messenger Astrophysics) project, a twin-satellite mission that is being developed for launch by NASA.

The promise of POEMMA is to open two cosmic windows. One will be the search for neutrinos above 20 PeV by looking for showers generated as tau-neutrinos exit the earth. The other is the study of UHECR above 30 EeV. POEMMA is planned to have an exposure one order of magnitude larger than achieved hitherto, with the prospect of rewriting '*the textbook on the most extreme astrophysical accelerators and fundamental physics interactions above terrestrial accelerator energies*'. Launch is planned for the late 2020s.

A successful POEMMA mission will surely require a follow-up on the ground. I have likened this to the construction of LEP to make high-precision studies of the Z and W particles discovered at the SPS. To cover ~30,000 km$^2$ will require cheaper technologies: water-Cherenkov detectors will surely remain the workhorse of the installation, and can probably be made more cheaply, while an inexpensive fluorescence detector, FAST, is being developed [30].

This giant array should have the fluorescence detectors spaced sufficiently close to cover the whole area with a sensitivity to energies as low as 0.5 EeV. This would allow a search for two showers expected to arrive simultaneously as a consequence of photo-disintegration of a nucleus in the solar photon field. This idea was proposed by Zatsepin and developed further by Gerasimova and Zatsepin [31, 32] in the 1950s. Taking $^{12}$C as an example, the showers produced by a neutron and $^{11}$C would differ by a factor of 11 in energy and so, in principle, the mass of the primary could be found rather directly. There seems no prospect of making such an observation with particle detectors as the showers produced by the neutron, and even by the $^{11}$C nucleus, would require an impractically dense spacing to measure them. In the earliest considerations, the deflection of the charged fragments in the interplanetary magnetic field was overlooked, but was taken into account in later work [33, 34]. For an iron primary [33] the separation of the showers is generally too large even for the size of array envisaged here, but for lighter primaries it has been shown that separations of only a 100 km or so are expected [34].

The approach proposed is a combination of largely brute force and elegance, the latter lying in the prospect of a direct measurement of the primary mass. This would be a hugely significant advance: in addition to its implications for cosmic-ray research, it would open the prospect of testing models of hadronic interactions rather directly at energies well-beyond what is possible at the LHC.

## 7. Closing remarks

Cosmic-ray physics remains a fertile field and has also spawned many other disciplines including particle physics, carbon dating, gamma-ray astronomy, neutrino astronomy, thunderstorm studies and muon tomography. Until 1953, the study of what we now call 'particle physics' took place in cosmic rays with many elementary particles ($e^+$, $\mu^{+/-}$, $\pi^{+/-}$ and many of the 'strange particles') first observed using cloud chambers and nuclear emulsions. The contributions from cosmic rays to particle physics rapidly declined in the early 1950s with the coming of accelerators: at the ICRC in Bagnères de Bigorre in 1953 Leprince-Ringuet remarked "*We have to face the grave question: what is the future of cosmic rays? Should we continue to struggle for a few new results or would it be better to turn to the machines?* [35].

Until relatively recently, the relationship between the particle physics and cosmic ray communities has been somewhat cool. On the one hand there has been envy at the amount of funding available to accelerator science while at the same time the cosmic-ray community has been slow to embrace the statistical rigour developed by its off-spring, often trumpeting results that did not stand up to closer inspection. In 1999, in an invited article on the origins of elementary particle physics, written to mark the 70th anniversary of Reviews of Modern Physics [36], Val Fitch, who shared the Nobel Prize with Jim Cronin in 1980 for their discovery of CP-violation, made the following remarks about our field.

"*Those who became interested in cosmic rays tended to be rugged individualists, to be iconoclastic, and to march to the drummer in their own heads rather than some distant one. After all, this was the period when nuclear physics was coming into its own, it was the fashionable subject, it was the subject that had the attention of the theorists, it was the subject for which great accelerators were being built. The cosmic-ray explorers eschewed all that and found their satisfactions in what might be called the backwater of the time.*"

I find these remarks surprising, particularly as Fitch and Cronin were close friends and Fitch was surely well-aware of Cronin's move to cosmic ray physics some ten years earlier. I do not think that he would now see our field as a backwater. The field is more vibrant that it has ever been and I can only envy those young enough to enjoy it.

## Acknowledgements

I am most grateful to the members of the Cosmic Ray Commission and the Local Organising Committee for the invitation to give the Hess Memorial Lecture. I would also like to thank Jaime Alvarez-Muñiz, Toshihiro Fujii, Stefan Funk, Marco Muzio, Angela Olinto, Jay Perrett and Michael Unger who provided, sometimes at short notice, diagrams and information that I was pleased to use in both my oral and/or written presentations. I would also grateful to Valerie Higgins, the archivist at FermiLab, whose efforts led to the discovery of the words of John Linsley shown in figure 5. This paper is not intended as a review and I apologise to those whose projects have been discussed only briefly or not at all. Rather it is an attempt to register some of the issues discussed in the public lecture that can be found on the Conference Web Site (https://www.icrc2019.org).

## References

[1] T C Weekes et al. *Astrophysical Journal* 342 379 1989
[2] HESS Collaboration *Nature* 531 476 2016
[3] R Mirzoyan *Highlight Talk at Madison ICRC 2019* POS(ICRC2019) 010
[4] HAWC Collaboration *arXiv:1905.1251*
[5] Tibet AS$\gamma$ Collaboration *arXiv:1906.0552*
[6] M G Artsen et al. *Physical Review Letters* 113 101101 2014
[7] A Castellina *Highlight Talk at Madison ICRC 2019* POS(ICRC2019) 004
[8] Auger Collaboration arXiv:1906.07422
[9] D Williams *Highlight Talk at Madison ICRC 2019* POS(ICRC2019) 016
[10] Auger Collaboration *Science* 357 1366 2017
[11] F Halzen *Review Talk at Madison ICRC 2019* POS(ICRC2019) 021
[12] M Unger, G R Farrar and L Anchordoqui *Phys Rev D* 92 123001 2015
[13] M S Muzio, M Unger and G R Farrar *arXiv:1906.06233*
[14] LHAASO Collaboration POS(ICRC2019) 217
[15] Lake Baikal GVD-Collaboration POS(ICRC2019) 873 and arXiv:1908.05458
[16] GEN2 Collaboration POS(ICRC2019) 1031
[17] KM3Net Collaboration arXiv:1601.07459
[18] S Ogio *Highlight Talk* at *Madison ICRC 2019* POS(ICRC2019) 013
[19] R Vazquez, F Halzen and E Zas *Phys Rev D* 45 356 1992
[20] P T Reynolds ICRC Dublin1991 OG 4.7-11 p 496
[21] J C Perrett and J T P M van Stekelenborg *J Phys G* 15 1291 1991
[22] K Greisen ICRC London 1965 Vol 2 p 609
[23] J V Jelley et al. *Nature* 205 3271965
[24] E Zas, F Halzen and T Stanev *Phys Rev D* 45 362 1992
[25] K Suga *Proc. 5th Interamerican Seminar on Cosmic Rays, La Paz, Bolivia*, Vol. 2
[26] T Hara et al. *Acta Phys. Acad. Sci. Hungaricae* 29 369 1970
[27] H E Bergeson et al. *Physical Review Letters* 39 847 1977
[28] C B A McCusker, H D Rathgeber and M M Winn ICRC Jaipur 1963 4 306
[29] M A Markov *Proc. 10th ICHEP, Rochester* p 578 1960
[30] T Fuji et al. POS(ICRC2019) 259
[31] G T Zatsepin *Dokl Akad Nauk SSSR* 80 577 1951
[32] N M Gerasimova and G T Zatsepin *Soviet Physics* 11 899 1960
[33] G Medina Tanco and A A Watson *Astroparticle Physics* 10 157 1999
[34] L Epele, S Mollerach and E Roulet *JHEP* 9903.017 1999
[35] J W Cronin *Eur Phys J H* 30 183 2011
[36] V Fitch *Rev Mod Phys* 70 S25 1999
[37] H Schoorlemmer for the SWGO Collaboration arXiv:1908.98858